# Fano resonance in Raman scattering of graphene

Duhee Yoon [a,†], Dongchan Jeong [b], Hu-Jong Lee [b], Riichiro Saito [c], Young-Woo Son [d], Hyun Cheol Lee [a], Hyeonsik Cheong [a,*]

[a] Department of Physics, Sogang University, Seoul 121-742, Korea,
[b] Department of Physics, Pohang University of Science and Technology, Pohang 790-784, Korea
[c] Department of Physics, Tohoku University, Sendai 980-8578, Japan,
[d] Korea Institute for Advanced Study, Seoul 130-722, Korea.

**ABSTRACT**

Fano resonances and their strong doping dependence are observed in Raman scattering of single-layer graphene (SLG). As the Fermi level is varied by a back-gate bias, the Raman G band of SLG exhibits an asymmetric line shape near the charge neutrality point as a manifestation of a Fano resonance, whereas the line shape is symmetric when the graphene sample is electron or hole doped. However, the G band of bilayer graphene (BLG) does not exhibit any Fano resonance regardless of doping. The observed Fano resonance can be interpreted as interferences between the phonon and excitonic many-body spectra in SLG. The absence of a Fano resonance in the Raman G band of BLG can be explained in the same framework since excitonic interactions are not expected in BLG.

[*] *Corresponding author:* Fax: +82 2 717 8434
   E-mail address: hcheong@sogang.ac.kr (H. Cheong)
[†] Current Address: Department of Engineering, University of Cambridge, Cambridge CB3 0FA, UK

# 1. Introduction

Fano resonances in light scattering from materials result from interferences between a discrete level and a continuum spectrum of elementary excitations [1–13]. Fano resonances in various spectra are characterized by asymmetric line broadening, expressed by the Breit-Wigner-Fano (BWF) line shape [2,3]. The BWF line shapes have been observed in Raman spectra of carbonic materials such as metallic single wall carbon nanotubes (m-SWCNTs) and graphite intercalation compounds [4–10]. In the case of m-SWCNTs, the BWF line shape has been observed in the G band when the Fermi energy $E_F$ satisfies $|E_F| < \frac{1}{2}\hbar\omega_G$, where $\omega_G$ is the frequency of the G band phonon [7–10]. In the case of graphene, Fano resonances have been observed in infrared spectra of bilayer graphene (BLG) in the presence of an energy band gap generated by a perpendicular electric field, which can be interpreted as the inference between the phonon and a continuum spectrum of electronic transitions near the bandgap [11,12]. Fano resonances in the low-frequency shear phonon modes, so-called *C* modes, of 3-layer or thicker graphene have also been observed in Raman spectra and interpreted as the interference between the *C* mode phonons and the continuum of the electronic transitions [13]. Dependence of Fano resonances in the infrared spectra upon the layer thickness and stacking order has also been reported for few layer graphene [14]. However, Fano resonances in Raman scattering of single-layer graphene (SLG) have not yet been observed. It turns out that the asymmetry factor for SLG is smaller than that of m-SWCNTs, making it difficult to observe and analyze BWF line shapes in SLG. Since Fano resonances do exist in Raman spectroscopy of m-SWCNTs [5–10], it is reasonable to expect that the BWF line shape should be observable in the Raman G band of SLG, which would provide insight on many-body excitations in graphene. In this paper, we report the observation of Fano resonances in the Raman spectrum of SLG as a function of the Fermi energy. The Fermi energy dependence of the asymmetry factor of the BWF line can be explained by the interference between the phonon and many-body excitonic excitations of massless Dirac fermions. Furthermore, we have also found that a BWF line shape does not appear in the G band of BLG regardless of doping.

Raman spectroscopy is one of the most widely used measurement techniques in graphene research. In addition to being one of the most straightforward methods to identify the number of graphene layers [15], Raman spectroscopy has been found useful in studying doping, strain, and mechanical properties [16–23]. Due to the interaction between zone-center phonons with electron-hole excitations, the optical phonon dispersion in graphene has a cusp structure at the center ($\Gamma$) or at the hexagonal corners ($K$ and $K'$) of the Brillouin zone, called the Kohn anomaly, exhibiting a linear phonon dispersion $\omega(k)$ when the wavevector $k$ is measured from the $\Gamma$ (or $K$) points [17–19]. When the Fermi energy shifts from the Dirac point by doping, the Kohn anomaly effect becomes suppressed and phonon hardening occurs [17–19]. Narrowing of the spectral width as well as the blue-shift of the G band phonon of SLG as a function of doping levels has been reported [18,19]. Although the Raman spectrum has important information on the interaction of phonons with many-body effects in SLG, the effect of doping on the Raman spectral line shape, especially on the asymmetry factor, has been neither observed nor discussed. Here we report observations of doping-dependent asymmetric line broadening of SLG and discuss its origin, contrasting it with the lack of such asymmetric line broadening in BLG.

## 2. Experimental

Graphene sheets used in this study were prepared on highly doped *n*-type silicon substrates capped with a 300-nm-thick $SiO_2$ layer by mechanical exfoliation from natural graphite flakes. SLG and BLG samples were identified by an optical microscope and Raman spectroscopy. The carrier density was controlled by the back-gate voltage. Electrical contacts were made by conventional e-beam lithography and e-beam evaporation of Ti/Au. Fig. 1(a) shows optical microscope images of the samples used in this study. The first sample had a graphene piece with both a SLG (S1) and a BLG (B1) region. We studied S1 and B1 independently by placing an electrode on the boundary between the two regions. The second sample consisted of only SLG (S2). The samples were mounted inside a microscope cryostat with either liquid nitrogen or helium cryogen (Oxford Instruments, HiResII). While a gate voltage ($V_g$) was applied to the highly doped *n*-type silicon substrate, Raman measurements were performed at 6, 100, and 300 K in vacuum as illustrated in Fig. 1(b). The 514.5-nm line of an Ar ion laser with a power of 0.3 mW was used as the excitation source. A 40× long-working-distance microscope objective (0.6 N.A.) was used to focus the laser beam onto the sample and collect the scattered light. The scattered light signal was dispersed by a Jobin-Yvon Triax 550 spectrometer (1800 grooves/mm) and detected with a liquid-nitrogen-cooled CCD detector. The spectral resolution was ~0.7 cm$^{-1}$.

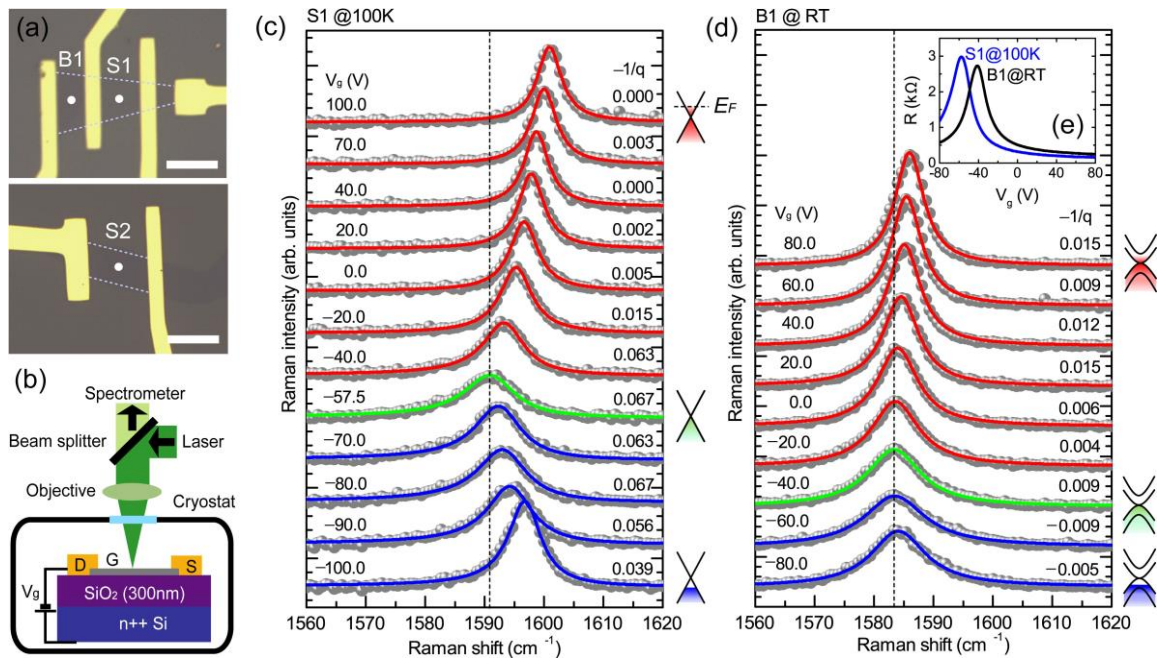

Fig. 1 – (a) Optical microscope images of the back-gated graphene samples. S1 and S2 are SLG samples, and B1 is a BLG sample. The scale bar is 10 μm. (b) Schematic of the low temperature Raman and transport measurements. (G: graphene, S: source, D: drain, $V_g$: back-gate bias voltage). (c) G band of the SLG sample (S1) as a function of the back-gate voltage ($V_g$) at 100 K. The Fermi energy $E_F$ is controlled by $V_g$. $-1/q$, characterizing the coupling strength between the phonon and continuum states, is obtained by fitting to the BWF line shape. (d) G band of the BLG (B1) as a function of $V_g$ at room temperature. (e) Source-drain resistance $R$ as a function of $V_g$.

**3. Results and discussion**

Figs. 1(c) and (d) present the variations of the Raman G band of samples S1 and B1 as a function of the gate voltage, measured at 100 K and at room temperature, respectively. The charge neutrality point, where the Fermi energy is at the Dirac point, in terms of $V_g$ was determined by measuring the source-drain resistance $R$ as a function of $V_g$ as shown in Fig. 1(e). The drift of the charge neutrality point during the measurements was negligible. In the case of sample S1 (B1), the spectrum for $V_g = -57.5$ V (–40 V) corresponds to the charge neutrality point. The Raman G band blue-shifts and becomes narrower when the gate voltage increases or decreases from the charge neutrality point, consistent with previous studies [18,19]. In Figs. 2(a) and (b), the peak position and the spectral width of the G band are plotted as a function of $E_F$. The back-gate voltage $V_g$ is converted to $E_F$ following a conversion formula reported in previous studies for SLG [24] and BLG [25]. The doping dependent shifts of Raman frequencies and the variation of the spectral line widths in Figs. 2(a) and (b) are consistent with previous reports [18,19], in which they were explained in terms of the Kohn anomaly suppression and the breakdown of the Born-Oppenheimer approximation. Due to unintentional doping by adsorbates from the environment, it was rather difficult to obtain stable hole doping (negative Fermi energy), especially at room temperature.

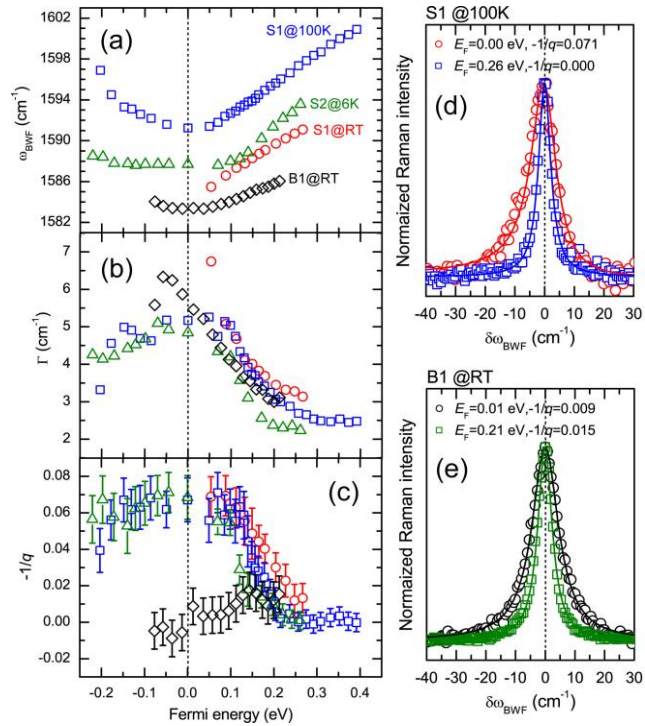

**Fig. 2 – (a) G band frequencies ($\omega_{BWF}$) of SLG and BLG samples as functions of the Fermi energy ($E_F$) at several temperatures ($T$). (b) Broadening parameter $\Gamma$ and (c) asymmetry factor $-1/q$, characterizing the coupling strength between the phonon and continuum states, as functions of $E_F$. (d) Comparison of the line shapes of the G band for doped (blue dots) and undoped (red dots) cases from a SLG sample (S1). The red and blue lines are obtained by fitting data to the BWF line shape of Eq. (1). (e) Same comparison for undoped (black) and electron doped (green) BLG sample (B1).**

When we focus our attention on the Raman spectral line shape of the G band, we can find an asymmetric line shape which deviates from a single Lorentzian. Near the charge neutrality point, the spectra are asymmetrically broadened at the lower-frequency side. In Fig. 2(d), two spectra from the SLG sample S1 are compared. One (red) is measured at the neutral point ($E_F = 0$ eV) and the other (blue) at the electron-doped case ($E_F = +0.26$ eV). The asymmetric broadening is seen only in the $E_F=0$ spectrum. The asymmetric broadening was observed at all measured temperatures (6, 100, and 300 K), which means that the broadening of the Fermi distribution function is not essential for explaining the asymmetry. Fig. 2(e) also compares two spectra taken from the BLG sample B1 at the charge neutrality point (black) and at $E_F=+0.21$ eV (green). In this case, the asymmetric broadening did not appear for both the cases. High signal-to-noise ratios for the spectra were essential for unambiguous identifications of the asymmetric broadening, and so a long data acquisition time of 200 seconds was used for each spectrum. In previous studies of doping-dependent Raman spectroscopy of SLG [18,19], low signal-to-noise ratios or a low spectral resolution might have made it difficult to discern such an effect.

Asymmetric line shapes in Raman spectra may result from several different physical origins. For example, Casiraghi *et al.* reported that some of the SLG samples exhibited asymmetric line shapes due to an inhomogeneous charge distribution within the laser spot [26]. In general, such a broadening can occur on either side of the Raman peak since charge inhomogeneity should be random. However, near the charge neutrality point, such inhomogeneous broadening should occur only on the *higher-frequency side* of the Raman peak because it is bounded below by the G-phonon frequency for the perfectly undoped ($E_F=0$) case. This is contrary to what was observed in our case, where the asymmetric broadening at the charge neutrality point always occurs on the *lower-frequency side* of the Raman G band. Therefore, we can eliminate charge inhomogeneity-induced broadening as the explanation for our result. In another example, an asymmetric line broadening on the *lower-frequency side* of a Raman peak has been observed in resonance Raman scattering of a semiconductor where the incident or scattered photons had energies close to an optical transition that involved a localized state [27]. In that case, phonons with nonzero *k* around the zone center were involved in Raman scattering because of the resonance with the localized state. Often in semiconductors, the highest-frequency phonon branch has the maximum at the zone-center. Consequently, the phonons around the zone center should have lower frequencies than that of the zone-center phonon, yielding asymmetric broadening on the lower energy side near the resonance. In the case of graphene, the highest Raman active optical phonon branch, the longitudinal optical (LO) phonon branch, has the minimum at the zone center due to the Kohn anomaly [17–19]. In addition, the in-plane transverse optical phonon (iTO) branch can contribute to the disorder-activated Raman scattering [28]. This branch has an opposite dispersion, having a maximum at the zone center. However, the dispersion of the LO branch is steeper than that of the iTO branch [29,30]. As a result, any combined effect of the two branches due to disorder should be on the *higher* energy side of the G band like the D′ band around 1620 cm$^{-1}$, which is again contrary to the observation. Furthermore, the lack of the D band signal implies that the samples have high crystallinity, which again rules out the possibility of the disorder-activated asymmetric broadening as the origin of our observation.

We interpret the observed asymmetric line broadening near $E_F$=0 as being due to a Fano resonance. Near a Fano resonance, the asymmetric Raman spectrum can be expressed by the BWF line shape [1,4–10], defined by

$$I(\omega) = I_0 \frac{[1+(\omega-\omega_{\text{BWF}})/q\Gamma]^2}{1+[(\omega-\omega_{\text{BWF}})/\Gamma]^2}, \qquad (1)$$

where the asymmetry factor $-1/q$ characterizes the coupling strength between the phonon and continuum spectra, $\Gamma$ is a broadening parameter, and $\omega_{\text{BWF}}$ is the uncoupled BWF peak frequency [13]. When $-1/q$ is large, the asymmetric line broadening on the lower-frequency side is accompanied by a 'dip' in the spectrum on the higher frequency side, which is regarded as the 'hallmark' of a Fano resonance. However, in Raman spectra of carbonic materials, such a 'dip' is usually missing even when $-1/q$ is much larger than what is observed in SLG [4–9]. Therefore, identification of a Fano resonance in such cases inevitably relies on phenomenological theoretical model that can consistently explain the experimental data. In Fig. 2(d), we fitted the experimental data for $E_F = 0$ eV to the BWF line shape by the solid lines, and the fitting works well. Fig. 2(c) summarizes the $-1/q$ values thus obtained for the three samples at different temperatures. For the SLG samples, the $-1/q$ values show a maximum value near $E_F = 0$ at all the measurement temperatures, whereas that for the BLG sample remains constant within error bars. The maximum $-1/q$ value of ~0.07 is somewhat smaller than what has been observed in m-SWCNTs (~0.02 – 0.4) [5–9] or graphite intercalation compounds (~0.2 – 1.3) [4], which may be attributed to the smaller density of states at the Fermi energy.

The absence of a Fano resonance in the Raman G band of BLG is a surprise, since Fano resonances in IR spectroscopy of BLG have been observed [11,12]. In the case of IR measurements, the absence of a Fano resonance in SLG is natural because the $E_{2g}$ phonon modes of the Raman G band are IR inactive, whereas the $E_u$ phonon mode in BLG is IR active [11,12]. Because the $E_{2g}$ phonon mode of SLG is Raman active, one may try to interpret the Fano resonance in the Raman G band of SLG using a model similar to the case of IR spectroscopy of BLG. However, in that case, a Fano resonance should be observed in the Raman G band of BLG, too. Therefore, the absence of a Fano resonance in the Raman G band of BLG implies that the origin of the Fano resonance observed in Raman scattering is different from that found in IR spectroscopy.

In the IR spectroscopy studies of BLG, either the electron-hole excitations between the higher-energy hole band and the lower-energy electron band [11,12] or the excitons formed in the presence of the bandgap opening due to electric fields [12] were suggested as the continuum spectra responsible for the Fano resonance. In the case of Raman scattering in m-SWCNTs, on the other hand, plasmons were often quoted to be the continuum spectra responsible for Fano resonances [6–8]. Recently, electronic Raman scattering has been observed for m-SWCNTs and understood as inelastic scattering by a continuum of low-energy electron-hole pairs created across the graphene-like linear electronic subbands [10]. Interference between such electronic Raman scattering and the phonon may give rise to a Fano resonance, but there has been no further discussion on BWF lines.

In order to explain the asymmetric line broadening in SLG in term of a Fano resonance, one needs to identify a continuum spectrum in the energy range of the G band phonon (~200 meV). Any model should also be able to explain the Fano resonance in SLG *and* the lack of it in BLG in a consistent way. We propose that the interference occurs between the phonon and electronic continuum states that are both strongly renormalized by excitonic processes. Unlike the previous cases of Fano resonances in carbonic materials where the continuum state is made of either a continuum of single-particle excitations or collective excitations near the Fermi surface, the Fano resonance in this case is given by a continuum of many-body states driven by electron-hole interactions as explained below qualitatively. Hereafter, we provide a phenomenological description of the aforementioned physical process with the following simple order-of-magnitude estimation. The standard expression for the binding energy of a Wannier exciton is $E_{ex} \approx m_{eff} e^4 / 2\hbar^2 \epsilon^2$. If we introduce a dimensionless coupling constant $\beta = e^2 / \hbar v_F \epsilon$, we can write $E_{ex} \approx m_{eff} v_F^2 \beta^2 / 2$. In the tight-binding model, $m_{eff} v_F^2 \sim t$, where $t$ is the hopping energy of ~3 eV [31]. The critical coupling constant corresponding to an excitonic instability is $\beta_c \approx 0.5$ [32]. Thus we obtain $E_{ex} \approx 0.5 t \beta_c^2 \approx 400$ meV, which is close to the optical phonon energy of ~200 meV.

The Raman intensity in the presence of a Fano resonance is usually expressed by [1]

$$I(\omega) = \frac{\pi \rho(\omega)[T_e(\omega_G - \omega) - VT_p]^2}{[\omega_G - \omega + V^2 R(\omega)]^2 + [\pi V^2 \rho(\omega)]^2} \tag{2}$$

where $R(\omega)$ and $\rho(\omega)$ are, respectively, the real and imaginary parts of the unperturbed Green's function for the electronic continuum. $V$ is the electron-phonon coupling strength, and $\omega_G$ the unperturbed $G$-phonon frequency. In Eq. (2), the terms of $V^2 R(\omega) = \Delta$ and $\pi V^2 \rho(\omega) = \Gamma$ can be interpreted as the real and imaginary parts, respectively, of the phonon self-energy. $T_e$ is the matrix element of the conventional Raman process [Fig. 3(a)] whereas $T_p$ is the matrix element of a Raman process where a phonon couples with a photon in the first order [Fig. 3(b)]. Basko recently demonstrated [33] that this process, which is usually ignored in most Raman scattering analyses, plays an important role in the Raman G band scattering in graphene. Then Eq. (2) can be recast into Eq. (1) by defining $-1/q$ as follows:

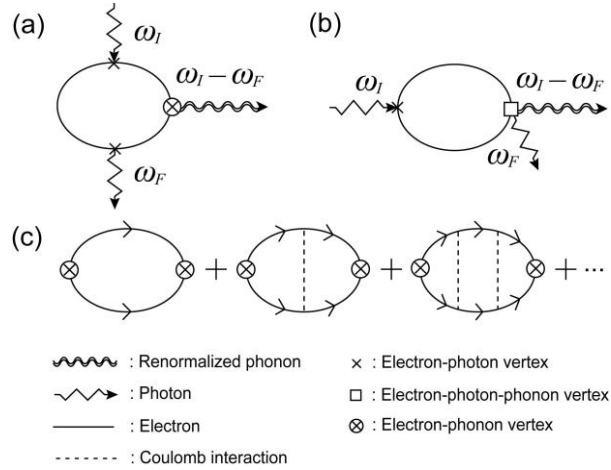

**Fig. 3 – Feynman diagrams for (a) the conventional Raman process where the incoming photon creates electron-hole excitation which in turn emits/absorbs a (renormalized) phonon and emits the outgoing photon sequentially. This corresponds to the $T_e$ term in Eq. (2); (b) Raman process where an incoming photon creates electron-hole excitation which emits/absorbs a (renormalized) phonon and the outgoing photon simultaneously. This corresponds to the $T_p$ term of Eq. (2); and (c) Excitonic processes contributing to the phonon self-energy of the G phonon.**

$$-\frac{1}{q} = -\frac{\Gamma}{VT_p/T_e + \Delta}. \quad (3)$$

The appearance of the asymmetric BWF line shape or the enhancement of $-1/q$ requires a sizable $T_p$ as discussed above, and it also needs the increase of the phonon self-energy ($\Delta$ and $\Gamma$) near the charge neutrality point. Recently, it was shown theoretically that excitonic many-body states form in the *undoped* limit of SLG [32,34]. We note that the phonon self-energy would increase by the formation of such excitonic many-body states because it is actually described by the same set of Feynman diagrams [Fig. 3(c)] as those contributing to the formation of excitonic many-body states [32]. Hence it follows that $\Delta$ and $\Gamma$ should increase in the *undoped* limit of SLG, leading to the increase of $-1/q$ as long as the first term in the denominator of Eq. (3) is not negligible. This would explain the observed BWF line shape in the undoped limit of SLG. Such excitonic processes would be suppressed in the *doped* case due to increased screening.

Furthermore, the absence of the Fano resonance in BLG can be explained in the same framework by comparing the (local) Green's function of BLG with that of metallic Fermi liquid. It is known [35,36] that the local Green's function of Fermi liquid in long time limit has the form

$$G_{\text{FL}}(t) \sim -\frac{D(E_F)}{t}, \quad (4)$$

where $D(E_F)$ is nonzero. It has been shown that the leading contribution of the local Green's function for BLG in the long time limit is given by [37]

$$G_{\text{bilayer}}(t) \sim -\frac{\gamma_1 - |E_F|}{8\pi v^2 t}, \quad (5)$$

where $\gamma_1$ is the interlayer coupling (~0.4 eV [31]) and $v$ the Fermi velocity. Here, even in the *undoped* limit ($E_F = 0$), the Green's function behaves like that of Fermi liquid, where no excitonic instability can be expected. This is a direct consequence of the interlayer coupling ($\gamma_1$). For $|E_F|$ values sufficiently smaller than $\gamma_1$, the numerator of Eq. (5) remains finite regardless of doping. Thus, one may conclude that a Fano resonance due to the interference between phonon excitations and excitonic many-body states cannot occur in BLG *at any doping level* used in our work. It should be mentioned that the absence of the Fano resonance in Raman scattering of BLG was also explained recently in the framework of a charged-phonon theory [38].

## 4. Conclusions

We have observed asymmetric line broadening of the Raman G band of SLG near the charge neutrality point, whereas such asymmetric line broadening is not observed for BLG, regardless of doping. We interpret the observed asymmetric line broadening as being due to a Fano resonance. We propose that the observed Fano resonance can be explained by interplay between the renormalized phonon excitation and excitonic processes in SLG. In the same framework, the lack of such asymmetric line broadening due to a Fano resonance in BLG can be explained because excitonic processes are suppressed in BLG.

**Acknowledgments**

This work was supported by the National Research Foundation (NRF) of Korea grant funded by the Ministry of Education, Science and Technology (MEST) of Korea (No. 2011-0017605, No. R17-2008-007-01001-0 and No. 2012-0008058) and a grant (No. 2011-0031630) from the Center for Advanced Soft Electronics under the Global Frontier Research Program of the MEST. Y.-W. S. was supported in part by the NRF grant funded by the MEST (QMMRC, No. R11-2008-053-01002-0). H.-J.L. was supported by NRF of Korea through SRC Center for Topological Matter (Grant No. 2011-0030788). R.S. acknowledges a MEXT grant (No. 20241023).